\documentclass{aastex}
\usepackage{spr-astr-addons}
\usepackage{url}

\RequirePackage{color}

\begin{document}

\title{Quasinormal Modes of  a quantum-corrected Schwarzschild black hole: gravitational and Dirac perturbations}
\slugcomment{Not to appear in Nonlearned J., 45.}
\shorttitle{QNMs of Quantum-Corrected Schwarzschild black hole}
\shortauthors{Mahamat S. et al.}

\author{Mahamat Saleh\altaffilmark{1}} \and \author{Bouetou Bouetou Thomas
        \altaffilmark{2,4,5}}
 \and
\author{Timoleon Crepin Kofane\altaffilmark{3,4,5,6}}

\email{mahsaleh2000@yahoo.fr}

\altaffiltext{1}{Department of Physics, Higher Teachers' Training
college, University of Maroua, P.O. Box 55 Maroua, Cameroon}
\altaffiltext{2}{Ecole Nationale Sup$\acute{e}$rieure
Polytechnique, University of Yaounde I, P.O. Box 8390, Cameroon}
\altaffiltext{3}{Department of Physics, Faculty of Science, University of
Yaounde I, P.O. Box 812, Cameroon}
\altaffiltext{4}{The African Center of Excellence in Information and Communication
Technologies (CETIC)}
\altaffiltext{5}{The Adus Salam International Center for Theoretical Physics, PO Box
586, Strada Costiera, IT-34014 Triestre, Italy}
\altaffiltext{6}{The Max Planck Institute for the Physics of Complex Systems,
N\"othnitzer Strasse 38, 01187 Dresden, Germany}

\begin{abstract}
In this work, quasinormal modes (QNMs)  of the Schwarzschild black hole are investigated by taking into account the quantum fluctuations. Gravitational and Dirac perturbations were considered for this case. The Regge-Wheeler gauge and the Dirac equation were used to derive the perturbation equations of the gravitational and Dirac fields respectively and  the third order Wentzel-Kramers-Brillouin (WKB) approximation method is used for the computing of the quasinormal frequencies. The results show that due to the quantum fluctuations in the background of the Schwarzschild black hole,
the QNMs of the black hole damp more slowly when increasing the quantum correction factor ($a$), and oscillate more slowly.
\end{abstract}


\section{Introduction}
\label{intro}
Theoretical studies on black holes have attracted substantial attention since the advent of General Relativity. It is well known that black holes can not be directly observed but they can be detected through their action on their neighborhoods. The behavior of the matter and fields surrounding a black hole not only tells us about its presence but also helps us determine its parameters. A real black hole can never be fully described by its basic parameters (mass, charge and angular momentum) and is always in the perturbed state (\cite{48}). For several decades, many efforts are made in investigating the evolution of external field perturbation around black holes. These perturbations lead to damped oscillations called quasinormal modes(\cite{49,22,23,24,25,26,27,28}). Different kinds of perturbation in the geometry of a black hole can excite certain combination of its characteristic frequencies of the normal modes for which the frequencies are no longer purely real showing that the system is losing energy. The real part of QN frequency represents the ring down frequency and
the imaginary part, the decay time. The resonant frequencies of the response of external perturbations called quasinormal frequencies are one of the most essential characteristics of black holes. It is widely believed that the QNMs carry a unique characteristic "fingerprint" which would lead to the direct identification of the black hole existence. Quasinormal modes associated with perturbations of different fields were considered in a lot of works (\cite{23,29,30,31,32,33,34,35,36,37,MelMos,38,39,40,konab,carkonlem,hod,mah,SW,ZER}).

We know that the vacuum undergoes quantum fluctuations. This phenomenon is the appearance of energetic particles out of nothing, as allowed by the uncertainty principle. Vacuum fluctuations have observable consequences like Casimir force between two plates in vacuum. Due to quantum fluctuations, the evolution of slices in the black hole geometry will lead to the creation of particle pairs which facilitates black holes' radiation (\cite{mathur}). It was shown that quantum vacuum fluctuations modify the geometry of the Schwarzschild black hole. \cite{kazsol} obtained new expressions of the Schwarzschild metric when the back reaction of the spacetime due to quantum fluctuations is taken into account.

Recently, Wontae and Yongwan (\cite{wontae}) investigated phase transition of the quantum-corrected Schwarzschild black hole and concluded that there appear a type of Grass-Perry-Yaffe phase transition due to the quantum vacuum fluctuations and this held even for the very small size black hole. More recently, we investigate QNMs of scalar perturbation around a quantum-corrected Schwarzschild black hole and show that the scalar field damps more slowly and oscillates more slowly due to the quantum fluctuations (\cite{ms}). In this paper, the QNMs of a
quantum-corrected Schwarzschild black hole is investigated in order to highlight the behavior of a gravitational and Dirac perturbations when the vacuum fluctuations are taken into account.

The paper is organized as follows. In section ~\ref{sec:1}, the quantum-corrected Schwarzschild black hole is presented. In section ~\ref{sec:01}, we derive the wave equation of a gravitational perturbation in the quantum-corrected Schwarzschild black hole background. In section ~\ref{sec:2}, the wave equation for Dirac perturbation of the black hole is derived. In section ~\ref{sec:3}, we evaluate the QN frequencies of the gravitational and Dirac perturbations by using the third
order WKB approximation method. The last section is devoted to a summary and
conclusion.

\section{The quantum-corrected Schwarzschild black hole}
\label{sec:1}
According to the work of Kazakov and Solodukhin on quantum deformation of the Schwarzschild
solution (\cite{kazsol}), the background metric of the Schwarzschild black hole is defined by
\begin{equation}\label{01}
    ds^2=f(r)dt^{2}-f(r)^{-1}dr^{2}-r^{2}(d\theta^{2}+sin^2\theta d\varphi^2),
\end{equation}
where
\begin{equation}\label{2}
    f(r)=-\frac{2M}{r}+\frac{1}{r}\int^r U(\rho)d\rho.
\end{equation}

For an empty space, $U(\rho)=1$. Thus we obtain the Schwarzschild
metric
\begin{equation}\label{3}
ds^{2}=\big(1-\frac{2M}{r}\big)dt^{2}-\big(1-\frac{2M}{r}\big)^{-1}dr^{2}-r^{2}(d\theta^{2}+sin^{2}\theta
d\varphi^{2}),
\end{equation}
where $M$ is the black hole mass.

Taking into account the quantum fluctuation of vacuum, the
quantity $U(\rho)$ transforms to (\cite{wontae})
\begin{equation}\label{4}
    U(\rho)=\frac{e^{-\rho}}{\sqrt{e^{-2\rho}-\frac{4}{\pi}G_R}},
\end{equation}
where $G_R=G_Nln(\mu/\mu_0)$, $G_N$ is the Newton constant and $\mu$ is a scale parameter.

The background metric of the quantum-corrected Schwarzschild black hole can then be read as

\begin{equation}\label{5}
ds^{2}=f(r)dt^{2}-f(r)^{-1}dr^{2}-r^{2}(d\theta^{2}+sin^{2}\theta
d\varphi^{2}),
\end{equation}
 with $f(r)=\big(-\frac{2M}{r}+\frac{\sqrt{r^{2}-a^{2}}}{r}\big)$ and $a^2=4G_R/\pi$.


The quantum correction parameter $a$ has the dimensionality of length $[l]$.

Using this form of the metric, the scalar curvature of the spacetime is transformed to
\begin{equation}\label{6}
    R=\frac{1}{a^2}\Bigg[\frac{2a^2}{r^2}\Bigg(1-\frac{1}{\sqrt{1-\frac{a^2}{r^2}}}\Bigg)+\frac{a^4}{r^4}\bigg(1-\frac{a^2}{r^2}\bigg)^{-3/2}\Bigg]
\end{equation}
 As we can see, the scalar curvature does not depend on the mass of the gravitating body but depends on the quantum correction parameter.

For large $r>>a$, the metric becomes
\begin{equation}\label{qc2}
    f(r)\simeq1-\frac{2M}{r}-\frac{a^2}{2r^2},
\end{equation}
which looks like the metric of a charged body with mass $M$ and imaginary charge $\pm i\frac{a}{\sqrt{2}}$. Thus, we can say that the quantum correction parameter $a$ acts as an imaginary charge added to the spacetime metric. Its coincides with the Reissner-Nordstr\"om metric but with an imaginary electric charge. It is therefore interesting for us to evaluate the QNMs of the Schwarzschild black hole when the quantum fluctuations modify it.

\section{Wave equation for gravitational perturbation of the black hole}
\label{sec:01}

For the gravitational perturbation, we will deal with the Regge-Wheeler gauge (\cite{23}). In this gauge, the gravitational perturbation is regarded as perturbation to the
background metric. We regard
the perturbed background metric $(\bar{g}_{\mu\nu})$ as the sum of the
unperturbed metric $(g_{\mu\nu})$ and the perturbation in it $(h_{\mu\nu})$
\begin{equation}\label{7}
    \bar{g}_{\mu\nu}=g_{\mu\nu}+h_{\mu\nu},
\end{equation}
where $\bar{g}_{\mu\nu}$ is the perturbed metric.

The perturbations $h_{\mu\nu}$ are supposed to be very small
compared with $g_{\mu\nu}$. The $h_{\mu\nu}$ can be calculated from
$g_{\mu\nu}$, and $R_{\mu\nu}+\delta R_{\mu\nu}$ from
$g_{\mu\nu}+h_{\mu\nu}$. The quantity $\delta R_{\mu\nu}$ can be
expressed in the form(\cite{Ei})

\begin{equation}\label{8}
    \delta{R}_{\mu\nu}=\delta\Gamma^{\alpha}_{\mu\alpha,\nu}-\delta\Gamma^{\alpha}_{\mu\nu,\alpha},
\end{equation}
where
\begin{equation}\label{9}
    \delta\Gamma^{k}_{\mu\nu}=\frac{1}{2}g^{k\alpha}\big(h_{\alpha\nu,\mu}+h_{\alpha\mu,\nu}-h_{\mu\nu,\alpha}\big).
\end{equation}
Similarly, we will use the unperturbed Christoffel symbols in
computing covariant derivatives of perturbation quantities. From
Eq. (\ref{8}) it follows that
\begin{equation}\label{10}
    \bar{g}^{\mu\nu}\simeq{g}^{\mu\nu}-h^{\mu\nu}.
\end{equation}
The perturbed Christoffel symbols are given by
\begin{equation}\label{11}
    \bar{\Gamma}^{k}_{\mu\nu}=\frac{1}{2}\bar{g}^{k\alpha}(\bar{g}_{\alpha\nu,\mu}+\bar{g}_{\alpha\mu,\nu}-\bar{g}_{\mu\nu,\alpha})=\Gamma^{k}_{\mu\nu}+\delta\Gamma^{k}_{\mu\nu},
\end{equation}
where
\begin{equation}\label{12}
    \delta\Gamma^{k}_{\mu\nu}=\frac{1}{2}g^{k\alpha}(h_{\alpha\nu,\mu}+h_{\alpha\mu,\nu}-h_{\mu\nu,\alpha}).
\end{equation}
The last equation reveals that the variation of the Christoffel
symbols, $\delta\Gamma^{k}_{\mu\nu}$, forms a tensor, even though
the Christoffel symbols themselves do not.


Adopting the \cite{23} matching, the final canonical form for an odd wave is then
\begin{equation}\label{13}
\begin{array}{lr}
    h_{\mu\nu}=&
\left(%
\begin{array}{cccc}
  0 & 0 & 0 & h_{0}(r) \\
  0 & 0 & 0 & h_{1}(r) \\
  0 & 0 & 0 & 0
  \\
  sym & sym & 0 & 0 \\
\end{array}\right)\exp(-ikT)\\%
&\times(sin\theta\partial_{\theta})P_{l}(\cos\theta).
\end{array}
\end{equation}

This form represents the perturbation of a spherically symmetric
black hole in Regge-Wheeler gauge.

The radial wave equation is the final equation which gives us the
behavior of the perturbation's oscillation. It can be reduced to a
simple linear differential equation.

\par\noindent In vacuum, the perturbed field equations simply reduce to (\cite{Nol})
\begin{equation}\label{14}
    \delta{R}_{\mu\nu}=0.
\end{equation}

\begin{figure}[h]
  \includegraphics[width=0.45\textwidth]{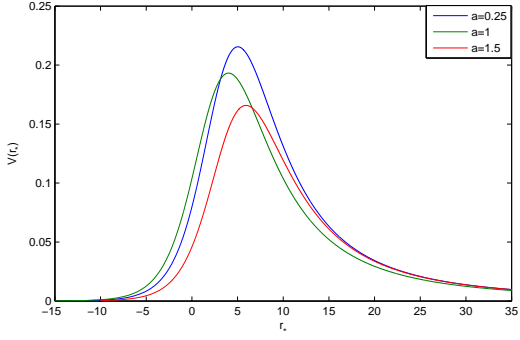}
\caption{Black hole's effective potential.}
\label{fig:1}       
\end{figure}

Using that and the background metric, we get from $\delta{R}_{23}=0$,
\begin{equation}\label{15}
\begin{array}{l}
    \Big(-\frac{2M}{r}+\frac{\sqrt{r^{2}-a^{2}}}{r}\Big)^{-1}i\omega{h}_{0}\\
    +\frac{d}{dr}\Big[\Big(-\frac{2M}{r}+\frac{\sqrt{r^{2}-a^{2}}}{r}\Big)h_{1}\Big]=0
\end{array}
\end{equation}
and from $\delta{R}_{13}=0$,
\begin{equation}\label{16}
\begin{array}{rr}
    \Big(-\frac{2M}{r}+\frac{\sqrt{r^{2}-a^{2}}}{r}\Big)^{-1}i\omega\Big(\frac{dh_{0}}{dr}+i\omega{h_{1}}-\frac{2h_{0}}{r}\Big)&\\
    +(l-1)(l+2)\frac{h_{1}}{r^{2}}&=0.
\end{array}
\end{equation}
Defining
\begin{equation}\label{17}
    \phi(r)=\Big(-\frac{2M}{r}+\frac{\sqrt{r^{2}-a^{2}}}{r}\Big)h_{1}/r
\end{equation}
and
\begin{equation}\label{18}
    \frac{dr_{*}}{dr}=\Big(-\frac{2M}{r}+\frac{\sqrt{r^{2}-a^{2}}}{r}\Big)^{-1}
\end{equation}
and eliminating $h_{0}$, we then get
\begin{equation}\label{19}
    \Big(\frac{d^{2}}{dr^{2}_{*}}+\omega^{2}\Big)\phi(r)=V\phi(r)
\end{equation}
where
\begin{equation}\label{20}
\begin{array}{ll}
    V=&\Big(-\frac{2M}{r}+\frac{\sqrt{r^{2}-a^{2}}}{r}\Big)\times\\
    &\Big(\frac{(l-1)(l+2)}{r^{2}}-\frac{6M}{r^{3}}+\frac{3\sqrt{r^{2}-a^{2}}}{r^3}-\frac{1}{r\sqrt{r^{2}-a^{2}}}\Big).
\end{array}
\end{equation}
Equation (\ref{19}) is a second order linear differential equation which
represents a perturbation equation. The potential $V$ depends
explicitly on the radius $r$. Its behavior is represented on Fig.\ref{fig:1}.

Through this figures, we can see that the potential decreases with increasing $a$. Moreover, the potential takes constant values when $r_*$ evolves to infinity. That allows us using the WKB approximation method  provided by Iyer and Will (\cite{SIW}) to evaluate the quasinormal frequencies.


\section{Wave equation for Dirac perturbation}
\label{sec:2}
For the general background spacetime, the massless Dirac equation is
\begin{equation}\label{dirac1}
    [\gamma^ae^\mu_a(\partial_\mu+\Gamma_\mu)]\Psi=0,
\end{equation}
where $\gamma^a$ is the Dirac matrix, $e^\mu_a$ is the inverse of the tetrad $e^a_\mu$ by the metric $g_{\mu\nu}=\eta_{ab}e^a_\mu e^b_\nu$, $\eta_{ab}$ is the Minkowski metric given by $\eta_{ab}=diag(-1,1,1,1)$, $\Gamma_\mu$ is the spin connection defined as $\Gamma_\mu=\frac{1}{8}[\gamma^a, \gamma^b]e^\nu_ae_{b\nu;\mu}$ and $e_{b\nu;\mu}=\partial_\mu e_{b\nu}-\Gamma^a_{\mu\nu}e_{ba}$.

In order to separate the Dirac equation, we must introduce the tetrad
\begin{equation}\label{dirac2}
    e^a_\mu=diag(\sqrt{f}, \frac{1}{\sqrt{f}}, r, rsin\theta).
\end{equation}
Using this expression, the Dirac equation becomes
\begin{equation}\label{dirac3}
\begin{array}{r}
    \frac{\gamma^0}{\sqrt{f}}\frac{\partial\psi}{\partial t}+\sqrt{f}\gamma^1(\frac{\partial}{\partial r}+\frac{1}{r}+\frac{1}{4f}\frac{df}{dr})\psi+\frac{\gamma^2}{r}(\frac{\partial}{\partial\theta}+\frac{1}{2}cot\theta)\psi\\
    +\frac{\gamma^3}{rsin\theta}\frac{\partial\psi}{\partial\varphi}=0.
\end{array}
\end{equation}
In order to simplify the above equation, we can define the perturbation $\psi$ as
\begin{equation}\label{dirac4}
    \psi=f^{-1/4}\phi.
\end{equation}
Then, Eq. (\ref{dirac3}) can be rewritten in its simplified form as
\begin{equation}\label{dirac5}
    \frac{\gamma^0}{\sqrt{f}}\frac{\partial\phi}{\partial t}+\sqrt{f}\gamma^1(\frac{\partial}{\partial r}+\frac{1}{r})\phi+\frac{\gamma^2}{r}(\frac{\partial}{\partial\theta}+\frac{1}{2}cot\theta)\phi+\frac{\gamma^3}{rsin\theta}\frac{\partial\phi}{\partial\varphi}=0.
\end{equation}
The Pauli matrices $\sigma^i$ are given by
\begin{equation}\label{dirac6}
    \sigma^1=\left(
               \begin{array}{cc}
                 0 & 1 \\
                 1 & 0 \\
               \end{array}
             \right); \sigma^2=\left(
               \begin{array}{cc}
                 0 & -i \\
                 i & 0 \\
               \end{array}
             \right); \sigma^3=\left(
               \begin{array}{cc}
                 1 & 0 \\
                 0 & -1 \\
               \end{array}
             \right).
\end{equation}
We introduce the tortoise coordinate $r_*=\int f^{-1}dr$ and the ansatz for the Dirac spinor
\begin{equation}\label{dirac7}
    \phi(t, r, \theta, \varphi)=\left(
                                  \begin{array}{c}
                                    \frac{iG^{(\pm)}(r)}{r}\chi^{\pm}_{jm}(\theta, \varphi) \\
                                    \frac{F^{(\pm)}(r)}{r}\chi^{\mp}_{jm}(\theta, \varphi) \\
                                  \end{array}
                                \right)e^{-i\omega t},
\end{equation}
where
\begin{equation}\label{dirac8}
    \chi^{+}_{jm}=\left(
                    \begin{array}{c}
                      \sqrt{\frac{j+m}{2j}Y^{m-1/2}_l} \\
                      \sqrt{\frac{j-m}{2j}Y^{m+1/2}_l} \\
                    \end{array}
                  \right) \ for \ j=l+\frac{1}{2},
\end{equation}
\begin{equation}\label{dirac9}
    \chi^{-}_{jm}=\left(
                    \begin{array}{c}
                      \sqrt{\frac{j+1-m}{2j+2}Y^{m-1/2}_l} \\
                      -\sqrt{\frac{j+1+m}{2j+2}Y^{m+1/2}_l} \\
                    \end{array}
                  \right) \ for \ j=l-\frac{1}{2},
\end{equation}
$Y^{m\pm1/2}_l(\theta, \varphi)$ represent the ordinary spherical harmonics.

\par
\begin{figure}
\begin{center}
  \includegraphics[width=0.45\textwidth]{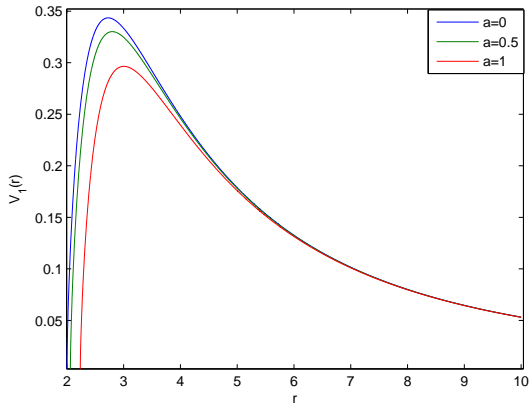}\\
  \caption{Behavior of the potential $V_1$ for $l=2$}\label{dp}
\end{center}
\end{figure}

\noindent Since
\begin{equation}\label{dirac10}
\begin{array}{l}
    \left(
      \begin{array}{cc}
        -i\big(\frac{\partial}{\partial\theta+\frac{1}{2}cot\theta}\big) & \frac{1}{sin\theta}\frac{\partial}{\partial\varphi} \\
        -\frac{1}{sin\theta}\frac{\partial}{\partial\varphi} & i\big(\frac{\partial}{\partial\theta+\frac{1}{2}cot\theta}\big) \\
      \end{array}
    \right)\left(
                   \begin{array}{c}
                     \chi^{\pm}_{jm} \\
                     \chi^{\mp}_{jm} \\
                   \end{array}
                 \right)\\
                 =i\left(
                            \begin{array}{cc}
                              k_{\pm} & 0 \\
                              0 & k_{\pm} \\
                            \end{array}
                          \right)\left(
                   \begin{array}{c}
                     \chi^{\pm}_{jm} \\
                     \chi^{\mp}_{jm} \\
                   \end{array}
                 \right),
\end{array}
\end{equation}
Equation (\ref{dirac3}) can be written in the simplified matrix form

\begin{equation}\label{dirac10}
\begin{array}{l}
    \left(
      \begin{array}{cc}
        0 & -\omega \\
        \omega & 0 \\
      \end{array}
    \right)\left(
                   \begin{array}{c}
                     F^{\pm} \\
                     G^{\pm} \\
                   \end{array}
                 \right)-\frac{\partial}{\partial r_*}\left(
                   \begin{array}{c}
                     F^{\pm} \\
                     G^{\pm} \\
                   \end{array}
                 \right)\\
                 +\sqrt{f}\left(
                            \begin{array}{cc}
                              \frac{k_{\pm}}{r} & 0 \\
                              0 & -\frac{k_{\pm}}{r} \\
                            \end{array}
                          \right)\left(
                   \begin{array}{c}
                     F^{\pm} \\
                     G^{\pm} \\
                   \end{array}
                 \right)=0.
\end{array}
\end{equation}
The cases (+) and (-) in the functions can be put together after some matching and the equation can be decoupled as (\cite{wang})
\begin{eqnarray}
  \frac{d^2F}{dr^2_*}+(\omega^2-V_1)F &=& 0, \\
  \frac{d^2G}{dr^2_*}+(\omega^2-V_2)G &=& 0,
\end{eqnarray}
with\begin{eqnarray}
       V_1 &=& \frac{\sqrt{f}\|k\|}{r^2}\Big(\|k\|\sqrt{f}+\frac{r}{2}\frac{df}{dr}-f\Big)\\
       \nonumber&& \texttt{for} \quad (k=j+\frac{1}{2},\ j=l+\frac{1}{2}), \\
       V_2 &=& \frac{\sqrt{f}\|k\|}{r^2}\Big(\|k\|\sqrt{f}-\frac{r}{2}\frac{df}{dr}+f\Big) \\
       \nonumber&& \texttt{for} \quad (k=-j-\frac{1}{2},\ j=l-\frac{1}{2}),
     \end{eqnarray}
where $k$ is the total angular momentum number.
The behavior of the potential is plotted on Fig.\ref{dp}.

\par\noindent These equations can be solved using the WKB approximation method to find the quasinormal modes.

\section{Quasinormal frequencies}
\label{sec:3} The wave equation (~\ref{19}) can be rewritten as:
\begin{equation}\label{21}
    \frac{d^{2}\psi}{dr^{2}_{*}}+Q(r_*)\psi=0,
\end{equation}
where $Q(r_*)=\omega^2-V(r_*)$.

For a black hole, the QN frequencies
correspond to solution of perturbation equation which satisfy the
boundary conditions appropriate for purely ingoing waves at the
horizon and purely outgoing waves at infinity. Incoming and
outgoing waves correspond to the radial solution proportional to
$e^{-i\omega r_{*}}$ and $e^{i\omega r_{*}}$, respectively. Only a
discrete set of complex frequencies satisfies these conditions.

To evaluate the QN frequencies, we applied here the third order WKB
approximation method derived by Schutz, Will (\cite{SW})and
Iyer (\cite{SIW}) to the above equation and these QN frequencies
are given by (\cite{yuzg2})
\begin{equation}\label{22}
    \omega^{2}=[V_{0}+(-2V''_{0})^{1/2}\tilde{\Lambda}]-i(n+\frac{1}{2})(-2V''_{0})^{1/2}[1+\tilde{\Omega}],
\end{equation}
where \begin{eqnarray}
      \nonumber
      \begin{tabular}{lll}
      $\tilde{\Lambda}$ &=& $\frac{1}{(-2V''_{0})^{1/2}}\bigg\{\frac{1}{8}(\frac{V^{(4)}_{0}}{V''_{0}})(\frac{1}{4}+\alpha^{2})$\\
      &&$-\frac{1}{288}(\frac{V'''_{0}}{V''_{0}})^{2}(7+60\alpha^{2})\bigg\}$ \\
     $ \tilde{\Omega}$
     &=&$\frac{1}{-2V''_{0}}\bigg\{\frac{5}{6912}(\frac{V'''_{0}}{V''_{0}})^{4}(77+188\alpha^{2})$\\
     && $-\frac{1}{384}(\frac{V'''^{2}_{0}V^{(4)}_{0}}{V''^{3}_{0}})(51+100\alpha^{2})$\\
    &&
     $-\frac{1}{288}(\frac{V^{(6)}_{0}}{V''_{0}})(5+4\alpha^{2})+\frac{1}{288}(\frac{V'''_{0}V^{(5)}_{0}}{V''^{2}_{0}})(19+28\alpha^{2})$\\
     &&$+\frac{1}{2304}(\frac{V^{(4)}_{0}}{V''_{0}})^{2}(67+68\alpha^{2})\bigg\}$,
     \end{tabular}
      \end{eqnarray}
      $\alpha=n+\frac{1}{2}$, and
      $V^{(n)}_{0}=\frac{d^{n}V}{dr^{n}_{*}}|_{r_{*}=r_{*}(r_{p})}$.

Using equation (~\ref{22}), we calculated numerically the QN frequencies of the gravitational perturbation for $M=1$  without and with  quantum correction of the black hole.  The results are shown in tables \ref{tab:1} and \ref{tab:2}, where $l$ is the
harmonic angular index, $n$ is the overtone number, $\omega$ is
the complex QN frequency and $a$ is the quantum correction parameter.

\begin{table*}[h!]
\small \caption{Quasinormal modes of gravitational perturbation
of Schwarzschild black hole }
\label{tab:1}       
\begin{tabular}{llllll}
\hline
$l$ & $\omega(n=0)$ & $\omega(n=1)$  & $\omega(n=2)$   & \multicolumn{1}{c}{$\omega(n=3)$} & $\omega(n=4)$ \\
\hline
2 & 0.3732 - 0.0892i & 0.3460 - 0.2749i & 0.3029 - 0.4711i & 0.2475 - 0.6729i &  \\
3 & 0.5993 - 0.0927i & 0.5824 - 0.2814i & 0.5532 - 0.4767i & 0.5157 - 0.6774i & 0.4711 - 0.8815i \\
4 & 0.8091 - 0.0942i & 0.7965 - 0.2844i & 0.7736 - 0.4790i & 0.7433 - 0.6783i & 0.7072 - 0.8813i \\
5 & 1.0123 - 0.0949i & 1.0021 - 0.2858i & 0.9833 - 0.4799i & 0.9575 - 0.6778i & 0.9264 - 0.8792i \\
\hline
\end{tabular}
\end{table*}
\begin{table*}[h!]
\small \caption{Quasinormal frequencies of gravitational perturbation
of quantum-corrected Schwarzschild black hole}
\label{tab:2}       
\begin{tabular}{lllllll}
\hline
$a$ & $l$ & $\omega(n=0)$ & $\omega(n=1)$  & $\omega(n=2)$   & \multicolumn{1}{c}{$\omega(n=3)$} & $\omega(n=4)$ \\
\hline
    & 2 & 0.3712 - 0.0891i & 0.3439 - 0.2745i & 0.3005 - 0.4704i & 0.2447 - 0.6720i & 0.1755 - 0.8776i \\
    & 3 & 0.5961 - 0.0926i & 0.5791 - 0.2809i & 0.5498 - 0.4759i & 0.5121 - 0.6764i & 0.4671 - 0.8802i \\
0.25 & 4 & 0.8049 - 0.0940i& 0.7922 - 0.2839i & 0.7692 - 0.4782i & 0.7387 - 0.6772i & 0.7024 - 0.8799i \\
    & 5 & 1.0070 - 0.0947i & 0.9968 - 0.2853i & 0.9778 - 0.4791i & 0.9519 - 0.6767i & 0.9206 - 0.8778i \\

\hline
    & 2 & 0.3654 - 0.0886i & 0.3377 - 0.2732i & 0.2936 - 0.4684i & 0.2368 - 0.6694i & 0.1663 - 0.8745i \\
    & 3 & 0.5870 - 0.0921i & 0.5697 - 0.2795i & 0.5399 - 0.4736i & 0.5016 - 0.6732i & 0.4558 - 0.8762i \\
0.5 & 4 & 0.7927 - 0.0935i & 0.7798 - 0.2824i & 0.7564 - 0.4758i & 0.7254 - 0.6739i & 0.6885 - 0.8758i \\
    & 5 & 0.9918 - 0.0942i & 0.9814 - 0.2838i & 0.9621 - 0.4767i & 0.9358 - 0.6734i & 0.9040 - 0.8736i \\

\hline
    & 2 & 0.3564 - 0.0879i & 0.3281 - 0.2712i & 0.2831 - 0.4652i & 0.2247 - 0.6651i & 0.1522 - 0.8694i \\
    & 3 & 0.5728 - 0.0912i & 0.5551 - 0.2771i & 0.5247 - 0.4698i & 0.4854 - 0.6680i & 0.4384 - 0.8698i \\
0.75& 4 & 0.7737 - 0.0927i & 0.7605 - 0.2800i & 0.7366 - 0.4719i & 0.7049 - 0.6686i & 0.6671 - 0.8691i \\
    & 5 & 0.9681 - 0.0934i & 0.9576 - 0.2814i & 0.9378 - 0.4727i & 0.9109 - 0.6680i & 0.8783 - 0.8668i \\

\hline
    & 2 & 0.3451 - 0.0869i & 0.3161 - 0.2684i & 0.2699 - 0.4608i & 0.2098 - 0.6592i & 0.1350 - 0.8623i \\
    & 3 & 0.5549 - 0.0901i & 0.5367 - 0.2739i & 0.5054 - 0.4647i & 0.4650 - 0.6611i & 0.4164 - 0.8610i \\
1   & 4 & 0.7497 - 0.0916i & 0.7361 - 0.2767i & 0.7116 - 0.4666i & 0.6790 - 0.6614i & 0.6400 - 0.8600i \\
    & 5 & 0.9382 - 0.0923i & 0.9273 - 0.2782i & 0.9070 - 0.4674i & 0.8793 - 0.6607i & 0.8458 - 0.8576i \\
\hline
    & 2 & 0.3322 - 0.0857i & 0.3025 - 0.2651i & 0.2552 - 0.4554i & 0.1933 - 0.6519i & 0.1163 - 0.8533i \\
    & 3 & 0.5346 - 0.0888i & 0.5158 - 0.2700i & 0.4836 - 0.4584i & 0.4420 - 0.6525i & 0.3918 - 0.8501i \\
1.25& 4 & 0.7224 - 0.0902i & 0.7084 - 0.2728i & 0.6832 - 0.4602i & 0.6496 - 0.6526i & 0.6095 - 0.8487i \\
    & 5 & 0.9042 - 0.0909i & 0.8930 - 0.2742i & 0.8721 - 0.4609i & 0.8436 - 0.6517i & 0.8091 - 0.8462i \\
\hline
\end{tabular}
\end{table*}
From these tables, we can remark that the real parts of the quasinormal frequencies as well as the absolute values of their imaginary parts of the quantum-corrected Schwarzschild black hole are smaller than those of the black hole without quantum correction. Moreover, these quantities decrease when increasing the quantum-correction parameter $a$.

For Dirac perturbation, the QN frequencies are given in tables \ref{tab:3} and \ref{tab:4}.
\begin{table*}[h!]
\small \caption{Quasinormal modes of Dirac perturbation
of Schwarzschild black hole }
\label{tab:3}       
\begin{tabular}{llllll}
\hline
$l$ & $\omega(n=0)$ & $\omega(n=1)$  & $\omega(n=2)$   & \multicolumn{1}{c}{$\omega(n=3)$} & $\omega(n=4)$ \\
\hline
2 & 0.5737-0.0963i & 0.5562-0.2930i & 0.5273-0.4972i & 0.4913-0.7068i & 0.4488-0.9192i \\
3 & 0.7672-0.0963i & 0.7540-0.2910i & 0.7304-0.4909i & 0.6999-0.6957i & 0.6640-0.9040i \\
4 & 0.9602-0.0963i & 0.9496-0.2902i & 0.9300-0.4876i & 0.9036-0.6892i & 0.8721-0.8944i \\
5 & 1.1530-0.0962i & 1.1442-0.2897i & 1.1274-0.4858i & 1.1043-0.6852i & 1.0761-0.8878i \\
\hline
\end{tabular}
\end{table*}
\begin{table*}[h!]
\small \caption{Quasinormal frequencies of Dirac perturbation
of quantum-corrected Schwarzschild black hole}
\label{tab:4}       
\begin{tabular}{lllllll}
\hline
$a$ & $l$ & $\omega(n=0)$ & $\omega(n=1)$  & $\omega(n=2)$   & \multicolumn{1}{c}{$\omega(n=3)$} & $\omega(n=4)$ \\
\hline
    & 2 & 0.5707 - 0.0962i & 0.5531 - 0.2925i & 0.5240 - 0.4965i & 0.4878 - 0.7058i & 0.4450 - 0.9180i\\
    & 3 & 0.7632 - 0.0961i & 0.7499 - 0.2906i & 0.7262 - 0.4901i & 0.6955 - 0.6947i & 0.6594 - 0.9027i\\
0.25& 4 & 0.9549 - 0.0961i & 0.9442 - 0.2898i & 0.9245 - 0.4870i & 0.8980 - 0.6883i & 0.8664 - 0.8933i\\
    & 5 & 1.1459 - 0.0962i & 1.1370 - 0.2895i & 1.1203 - 0.4854i & 1.0971 - 0.6846i & 1.0690 - 0.8871i\\

\hline
    & 2 & 0.5620 - 0.0957i & 0.5441 - 0.2911i & 0.5145 - 0.4943i & 0.4777 - 0.7029i & 0.4340 - 0.9143i\\
    & 3 & 0.7511 - 0.0957i & 0.7376 - 0.2893i & 0.7137 - 0.4881i & 0.6826 - 0.6919i & 0.6459 - 0.8991i\\
0.5 & 4 & 0.9388 - 0.0958i & 0.9281 - 0.2888i & 0.9083 - 0.4854i & 0.8816 - 0.6860i & 0.8496 - 0.8902i\\
    & 5 & 1.1258 - 0.0959i & 1.1169 - 0.2887i & 1.1001 - 0.4841i & 1.0769 - 0.6827i & 1.0487 - 0.8845i\\

\hline
    & 2 & 0.5469 - 0.0952i & 0.5288 - 0.2896i & 0.4989 - 0.4917i & 0.4614 - 0.6990i & 0.4168 - 0.9092i\\
    & 3 & 0.7335 - 0.0949i & 0.7196 - 0.2869i & 0.6950 - 0.4842i & 0.6631 - 0.6867i & 0.6254 - 0.8926i\\
0.75& 4 & 0.9186 - 0.0948i & 0.9074 - 0.2859i & 0.8869 - 0.4806i & 0.8592 - 0.6797i & 0.8262 - 0.8824i\\
    & 5 & 1.1031 - 0.0948i & 1.0938 - 0.2854i & 1.0762 - 0.4787i & 1.0520 - 0.6754i & 1.0226 - 0.8756i\\

\hline
    & 2 & 0.5315 - 0.0938i & 0.5125 - 0.2858i & 0.4813 - 0.4860i & 0.4424 - 0.6916i & 0.3960 - 0.9002i\\
    & 3 & 0.7112 - 0.0937i & 0.6969 - 0.2836i & 0.6715 - 0.4790i & 0.6386 - 0.6795i & 0.5997 - 0.8837i\\
1   & 4 & 0.8900 - 0.0937i & 0.8785 - 0.2828i & 0.8574 - 0.4756i & 0.8290 - 0.6728i & 0.7951 - 0.8736i\\
    & 5 & 1.0680 - 0.0938i & 1.0585 - 0.2824i & 1.0405 - 0.4738i & 1.0157 - 0.6688i & 0.9856 - 0.8671i\\
\hline
\end{tabular}
\end{table*}
As observed  for the gravitational perturbation, we can also remark that the real parts and the absolute values of the imaginary parts of the quasinormal frequencies decrease when increasing the quantum correction parameter. The particularity of Dirac perturbation here is that the imaginary parts of the ($n=0$) frequencies do not depend on the angular harmonic index ($l$).

\begin{figure}[h!]
\begin{center}
  \includegraphics[width=0.45\textwidth]{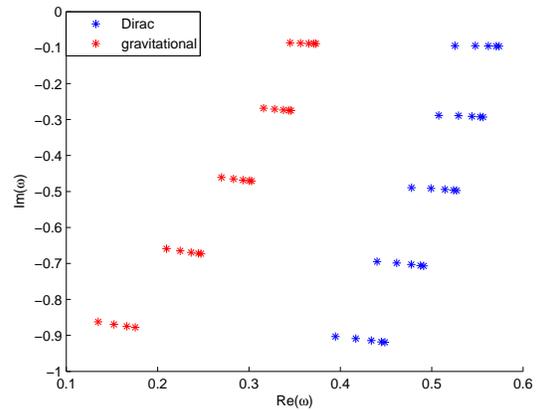}
\caption{Behavior of the quasinormal frequencies for $l=2$.}
\label{fig:2}       
\end{center}
\end{figure}

\begin{figure}[h!]
\begin{center}
  \includegraphics[width=0.45\textwidth]{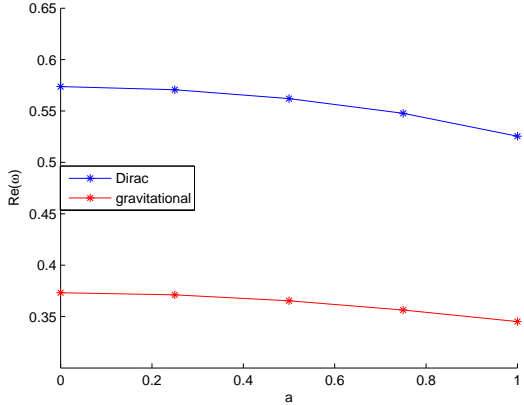}
\caption{Behavior of the real part of the quasinormal frequencies for $l=2$.}
\label{fig:02}       
\end{center}
\end{figure}

\begin{figure}[h!]
\begin{center}
  \includegraphics[width=0.45\textwidth]{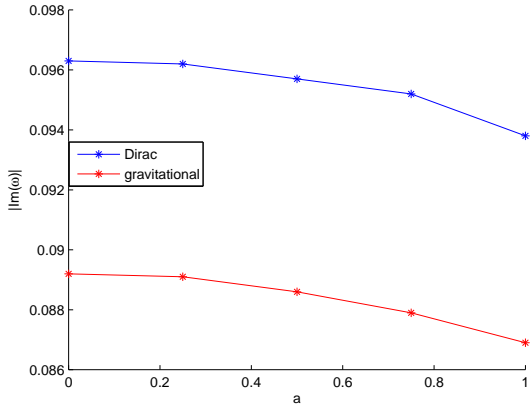}
\caption{Behavior of the imaginary part of the quasinormal frequencies for $l=2$.}
\label{fig:002}       
\end{center}
\end{figure}

The relationship between the real and imaginary parts of quasinormal frequencies of the gravitational as well as Dirac perturbations in
the background of the black hole is plotted in Fig.\ref{fig:2}. The behaviors of the real and imaginary parts of the quasinormal frequencies when varying the quantum-correction parameter, $a$, are also plotted (see Figs. \ref{fig:02} and \ref{fig:002}).

\section{Summary and Conclusion}
\label{conc} In summary, QNMs of a gravitational and Dirac perturbations
around a Schwarzschild black hole were evaluated using the
third order WKB approximation. For the gravitational perturbation, the results of table~\ref{tab:1}
are obtained without considering any kind of correction while those of
table~\ref{tab:2} are obtained when considering the quantum-corrected Schwarzschild black hole metric. Dirac quasinormal frequencies are summarized in tables \ref{tab:3} and \ref{tab:4} for the cases of the Schwarzschild black hole with and without quantum correction, respectively.
Through the above tables, we can remark that the absolute values
of the imaginary parts of the quasinormal frequencies of the quantum-corrected black hole are smaller compared to those without quantum correction,
for fixed set of $l$ and $n$. Moreover, we can remark that these values decrease when increasing $a$. That can be clearly seen through Figs. \ref{fig:2} and \ref{fig:002}. Through Fig. \ref{fig:2}, we an also remark that the real parts of Dirac quasinormal frequencies are higher compared to gravitational ones. That implies that Dirac QNMs oscillate more rapidly than gravitational ones. Concerning the real parts, they are also decreasing when increasing $a$. Thus, we can conclude that, due to the quantum fluctuations, the quasinormal modes of gravitational and Dirac perturbations of the Schwarzschild black hole damp more slowly and oscillate more slowly.

\end{document}